# Brownian Dynamics Simulation of Nucleosome Formation and Disruption under Stretching

Wei Li, Shuo-Xing Dou, Peng-Ye Wang[*]

*Laboratory of Soft Matter Physics, Institute of Physics, Chinese Academy of Sciences, Beijing 100080, China*

**Abstract**

Using a Brownian dynamics simulation, we numerically studied the interaction of DNA with histone and proposed an octamer-rotation model to describe the process of nucleosome formation. Nucleosome disruption under stretching was also simulated. The theoretical curves of extension versus time as well as of force versus extension are consistent with previous experimental results.

## 1. Introduction

In all eukaryotic genomes, the basic unit of chromatin structure is nucleosome which is made up of 146 bp of DNA and histones ― H2A, H2B, H3, H4 and H1 (H5) (Wolffe, 1998). Histones H2A, H2B, H3 and H4, with each contributing two molecules, form a histone octamer around which DNA wraps in approximate two turns in a left-handed way. H1 (H5) is a linker histone. It binds at the point where DNA enters and exits the nucleosome. The histone octamer is a tripartite assembly in which a centrally located $(H3-H4)_2$ tetramer is flanked by two H2A-H2B dimmers, and the eight histone molecules form a left-handed protein superhelix (Klug *et al.,* 1980; Arents *et al.,* 1991; Hamiche *et al.,* 1996; Luger *et al.,* 1997; Luger and Richmond, 1998). Nucleosome represents the first level of packing of DNA in chromatin in a nucleus. At a higher level, many nucleosomes form a 'zig-zag' structure (Woodcock *et al.,* 1993; Leuba *et al.,* 1994). As DNA is negatively charged and histones are positively charged, the main interaction between DNA and histones should be of electrostatic nature. With the development of single molecule manipulation methods, quite a few experiments have been done on stretching DNA molecules to investigate their mechanical behaviors (Smith *et al.,* 1992; Strick *et al.,* 1996; Cluzel *et al.,* 1996; Wang *et al.,* 1997; Katritch *et al.,* 2000; Cui and Bustamante, 2000). In the case of chromatin stretching with optical tweezers, force-induced nucleosome disruption was clearly observed (Bennink *et al.,* 2002; Brower-Toland *et al.,* 2002).

In parallel with experimental studies of DNA and chromatin, many numerical studies have been done on the structural formation of polymer chain systems by molecular dynamics simulations or by Monte Carlo simulations. (Stevens and Kremer, 1995; Noguchi *et al.,* 1996; Fujiwara and Sato, 1997; Noguchi and Yoshikawa, 1998; Chodanowski and Stoll, 1999). Some numerical studies of DNA have given good explanations to experimental results (Schellman and Harey, 1995; Marko and Siggia, 1995, 1997; Marky and Manning, 1995; Coleman *et al.*, 2000; Kunze and Netz, 2000). For kinetics studies of DNA or nucleosome, a Brownian dynamics simulation method has been developed, with which Noguchi and Yoshikawa (2000) studied the toroidal morphology of DNA chain, and Sakaue *et al.*, (2001) studied the Brownian motion of histone core particle along DNA. The previous numerical studies have demonstrated that Brownian dynamics simulation is a useful method for studying the kinetics of DNA and nucleosome. In our earlier work, we built a simple model to describe the interaction between DNA and a histone octamer (Li *et al.*, 2003). In the present paper, by taking into account the fact that at some locations on DNA the interaction between the histone octamer and DNA is stronger than elsewhere, we used Brownian dynamics simulations to study the process of nucleosome formation and that of disruption under stretching. The theoretical results on the mechanical behaviors of nucleosomal arrays under stretching are compared with that obtained experimentally by Brower-Toland *et al.*, (2002).

---

[*] Corresponding author. E-mail address: pywang@aphy.iphy.ac.cn

## 2. Model

The DNA chain is modeled as a semi-flexible homopolymer chain and a histone octamer as a spherical core particle. The homopolymer chain is further modeled by *N* spheres that are connected by bonds. As in the experiment of Brower-Toland *et al.*, (2002), H1 histone is absent and thus not taken into account.

The self-avoiding effect of DNA chain is considered by using the repulsive part of the Morse potential,

$$U_{m,rep} = \varepsilon_m k_B T \sum \exp\{-\alpha_m (r_{i,j} - \sigma_m)\} , \qquad (1)$$

where $\varepsilon_m = 0.2$ and $\alpha_m = 2.4$. In this potential, $k_B$ is the Boltzmann constant and we set it to unity in this paper. *T* is the absolute temperature of the system, and we choose 298 K for it. $r_{i,j}$ is the distance between the *i*th and *j*th spheres of the DNA chain. $\sigma_m$ is the minimum distance between two spheres of the DNA chain which gives the self-avoiding separation. Thus the diameter of one sphere of the DNA chain is $\sigma_m$. As the width of DNA molecules is 2.3 nm (i.e., $\sigma_m = 2.3$ nm), one sphere in our model corresponds to the length of 7 base pairs (bp) of DNA. Throughout the paper we use $\sigma_m$ as the length unit.

The bonds between neighboring spheres of the DNA chain are considered through harmonic bonding potential,

$$U_{bond} = \frac{k k_B T}{2\sigma_m^2} \sum (|\vec{r}_i - \vec{r}_{i+1}| - \sigma_m)^2 , \qquad (2)$$

where $k = 400$, $\vec{r}_i$ and $\vec{r}_{i+1}$ are the location vectors of the *i*th and (*i+1*)th spheres of the DNA

chain. We model the chain stiffness by using the bending potential,

$$U_{bend} = \kappa k_B T \sum (1 - \frac{(\vec{r}_{i-1} - \vec{r}_i) \cdot (\vec{r}_i - \vec{r}_{i+1})}{\sigma_m^2}) , \qquad (3)$$

where we choose $\kappa = 5$ so that the DNA's persistence length in our model is consistent with the actual value, ~50 nm at physiological salt concentration of 0.1 M (Kunze and Netz, 2000).

The interaction between the DNA chain and a histone octamer is simulated with the Morse potential,

$$U_M = \varepsilon k_B T \sum [\exp\{-2\alpha (r_i - \sigma)\} - 2\exp\{-\alpha (r_i - \sigma)\}] , \qquad (4a)$$

where $\varepsilon = 6$, $\alpha = 6$. $r_i$ is the distance between the histone octamer and the *i*th sphere of the DNA chain. The diameter of the histone octamer is chosen as ~$2.8\sigma_m$ so that the relative sizes of the histone octamer and DNA are the same as the actual ones: in nature, the width of DNA is about 2.3 nm and the diameter of a histone octamer is about 6.4 nm. Thus the equilibrium distance between a sphere and the histone octamer is $\sigma = 0.5\sigma_m + 1.4\sigma_m = 1.9\sigma_m$. In the interaction potential as expressed in Eq. (4a), a large value of parameter $\varepsilon$ corresponds to a low salt concentration in the experiment.

Mapping methods *in vivo* reveal that in some cases nucleosomes are preferentially localized at specific genomic positions. The positioning may prevent specific protein binding to nucleosomal DNA or facilitate binding proteins to recognize specific DNA sequences (Straka and Horz, 1991; Lu *et al.,* 1995). On the other hand, it has been shown that a histone octamer has some specific sites which strongly bind with DNA (Luger *et al.,* 1997; Luger and Richmond, 1998; Wolffe, 1998) To numerically simulate the sequence preference and the nonuniform binding, we assume some locations (termed S-locations) on DNA where the interaction of DNA with the histone octamer is much stronger than elsewhere (see Fig. 1).

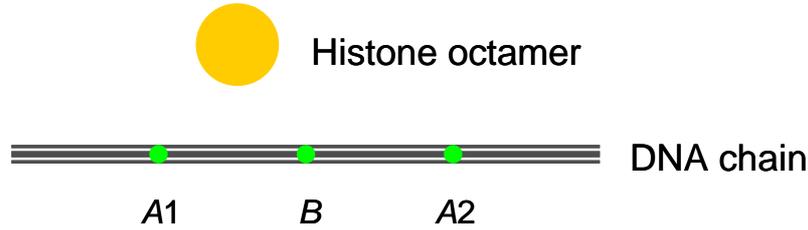

FIG 1. Schematic illustration of the histone octamer and DNA with three equally-distanced S-locations (green): *A*1, *A*2 and *B*.

The interactions between the histone octamer and the three S-locations are chosen as follow,

$$U_{M,A1} = \varepsilon' k_B T \sum [\exp\{-2\alpha(r_{A1} - \sigma)\} - 2\exp\{-\alpha(r_{A1} - \sigma)\}], \quad \text{S-location } A1 \quad (4b)$$

$$U_{M,A2} = \varepsilon' k_B T \sum [\exp\{-2\alpha(r_{A2} - \sigma)\} - 2\exp\{-\alpha(r_{A2} - \sigma)\}], \quad \text{S-location } A2 \quad (4c)$$

$$U_{M,B} = \varepsilon'' k_B T \sum [\exp\{-2\alpha(r_B - \sigma)\} - 2\exp\{-\alpha(r_B - \sigma)\}], \quad \text{S-location } B \quad (4d)$$

where $\varepsilon' = 15\varepsilon$ and $\varepsilon'' = 3\varepsilon$. $r_{A1}$, $r_{A2}$ and $r_B$ are the distances between the histone octamer and S-locations $A1$、$A2$ and $B$, respectively. As will be discussed later, the separations of these S-locations are closely related to the stretching curves obtained.

In the case of more than one histone octamer, we use the repulsive part of the Morse potential for the coulomb repulsion among them,

$$U_{M,rep} = \varepsilon_M k_B T \sum \exp\{-\alpha_M (R_{i,j} - \sigma_M)\}, \quad (5)$$

where $\varepsilon_M = 5$, $\alpha_M = 6$. $\sigma_M$ is the minimum distance between two neighboring histone octamers and is chosen as $\sigma_M = 3\sigma_m$. $R_{i,j}$ is the distance between the $i$th and $j$th histone octamers.

The overdamped Langevin equations are used to describe the motion of each sphere of the DNA chain and each histone octamer,

$$-\gamma_m \frac{d\vec{r}_i}{dt} + \vec{R}_{m,i}(t) - \frac{\partial U}{\partial \vec{r}_i} = 0, \quad (6a)$$

$$-\gamma_M \frac{d\vec{R}_j}{dt} + \vec{R}_{M,j}(t) - \frac{\partial U}{\partial \vec{R}_j} = 0, \quad (6b)$$

where $\gamma_m$ and $\gamma_M$ are the friction constants of a sphere and a histone octamer, respectively. They are calculated according to Stokes law. $\vec{R}_{m,i}$ and $\vec{R}_{M,j}$ are the Gaussian white noises which obey the fluctuation-dissipation theorem,

$$<\vec{R}_{m,i}(t)> = 0, \ <\vec{R}_{m,i}(t) \cdot \vec{R}_{m,j}(t')> = 6k_B T \gamma_m \delta_{i,j} \delta(t-t'), \quad (7a)$$

$$<\vec{R}_{M,i}(t)> = 0, \ <\vec{R}_{M,i}(t) \cdot \vec{R}_{M,j}(t')> = 6k_B T \gamma_M \delta_{i,j} \delta(t-t'). \quad (7b)$$

The total internal energy $U$ in Eqs. (6a) and (6b) consists of five terms: $U = U_{m,rep} + U_{bond} + U_{bend} + U_M + U_{M,rep}$.

For the process of force-induced nucleosome disruption, we modify Eq. (6a) as following,

$$-\gamma_m \frac{d\vec{r}_i}{dt} + \vec{R}_{m,i}(t) - \frac{\partial U}{\partial \vec{r}_i} + \vec{F} = 0, \quad i = 1 \quad (8a)$$

$$-\gamma_m \frac{d\vec{r}_i}{dt} + \vec{R}_{m,i}(t) - \frac{\partial U}{\partial \vec{r}_i} - \vec{F} = 0, \quad i = N \quad (8b)$$

$$-\gamma_m \frac{d\vec{r}_i}{dt} + \vec{R}_{m,i}(t) - \frac{\partial U}{\partial \vec{r}_i} = 0 , \qquad i \neq 1, N \qquad (8c)$$

where $\vec{F}$ is the force exerted on the first sphere of the DNA chain and $-\vec{F}$ is the force exerted on the $N$th sphere (the last one) of the DNA chain.

We perform the dynamics of this system using a stochastic Runge-Kutta algorithm (white noise) (Honeycutt, 1992). We choose $k_B T$ as the unit energy, $\sigma_m$ as the unit length, and $\gamma_m \cdot \sigma_m / \sqrt{T}$ as the unit time step in our simulation.

## 3. Results

The dynamic process of interaction between a 40-sphere DNA (70-sphere DNA) and a histone octamer (two histone octamers) is shown in detail in Fig. 2 (Fig. 3). It can be seen that the dynamic processes are quite similar in the two cases. DNA and the histone octamer(s) first come close to each other, then DNA wraps around the octamer(s) gradually to about two turns, forming a stable nucleosome structure. It can be noticed in both cases that during DNA wrapping, the shape of the DNA chain nearby the histone octamer changes drastically while other parts of the DNA chain are only drawn towards the histone octamer. And the wrapping involves looping of the DNA chain and then rotation of the looped DNA over the surface of the spherical histone octamers [from (d) to (e) in the case of Fig. 2]. In the simulation the histone octamer does not rotate, but as will be discussed later, it should rotate simultaneously with the looped DNA. From Figs. 2 and 3 the importance of the S-locations on DNA in the nucleosome formation can also be clearly noticed: they determine both the wrapping process and nucleosome positioning. The relative positions of the three S-locations in all nucleosomes are the same and agree with that given by Brower-Toland *et al.*, (2002). Note that the separation between S-locations $A1$ and $A2$ used in the simulations is $12\,\sigma_m$, i.e., ~81 bp of DNA.

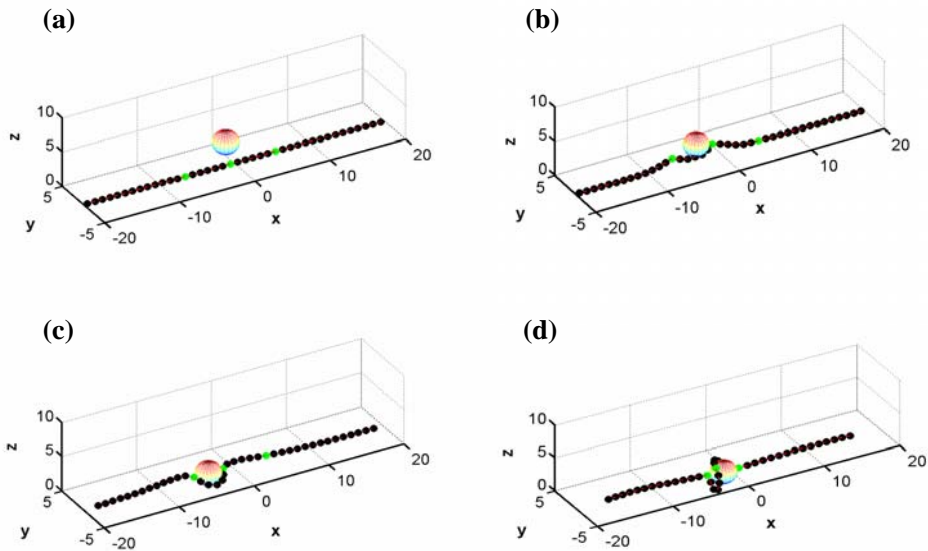

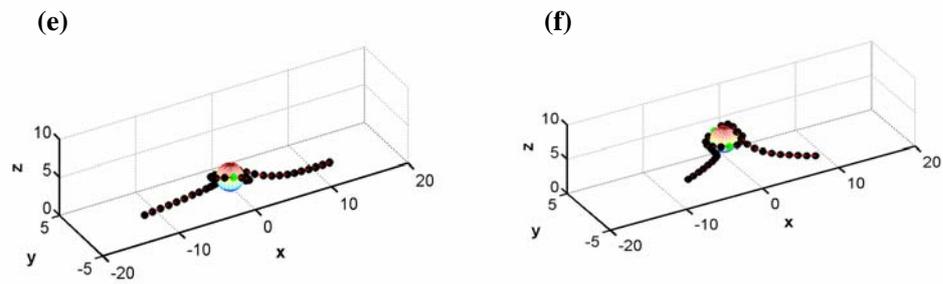

FIG 2. Snapshots of the dynamic process of interaction between a 40-sphere DNA and a histone octamer. The time step is $6\times 10^{-5}$ and the simulation time *t* is: (a) 0; (b) 0.3; (c) 4.5; (d) 9; (e) 45; (f) 90. The S-locations on DNA are shown in green here and in following figures.

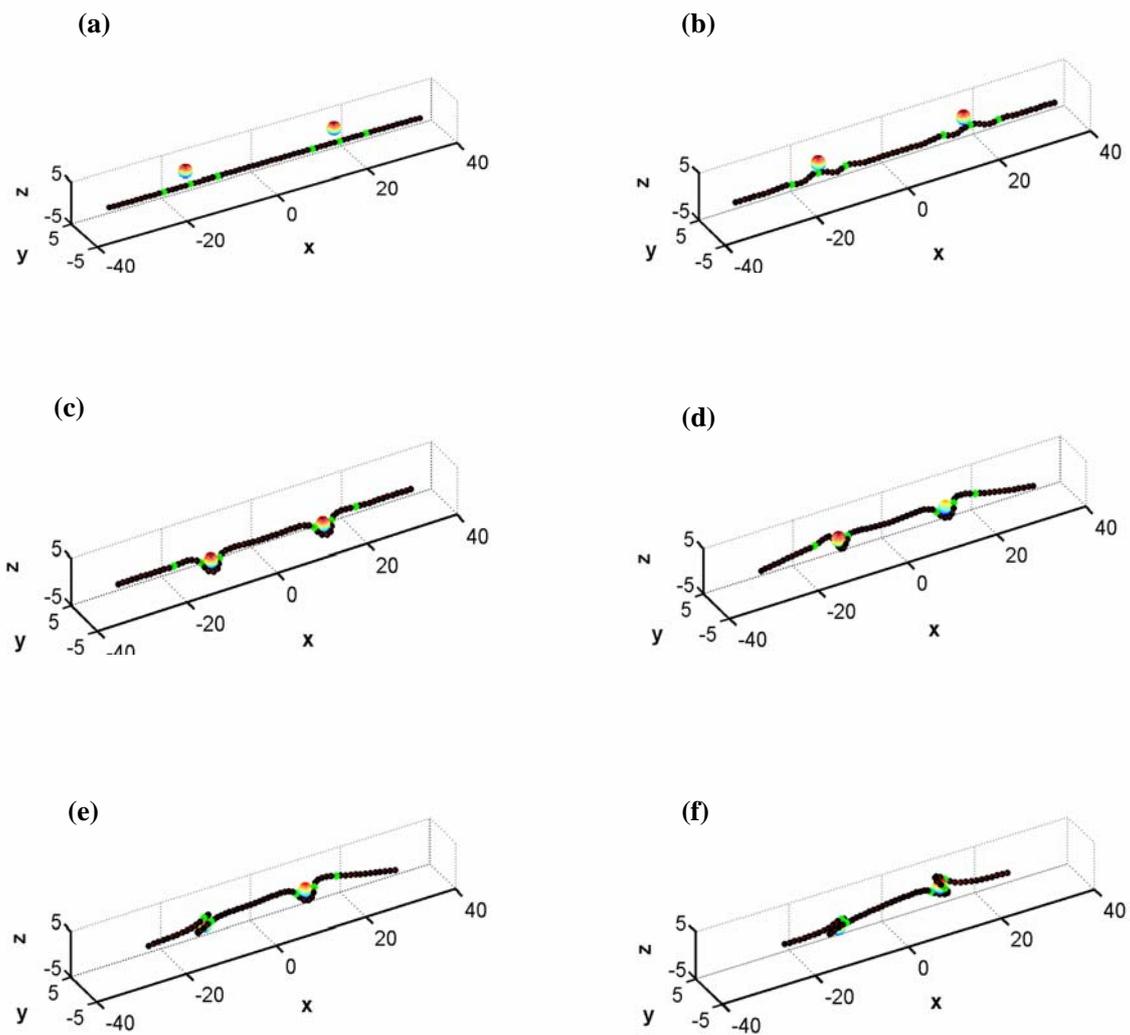

FIG 3. Snapshots of the dynamic process of interaction between a 70-sphere DNA and two histone octamers. The simulation time *t* is: (a) 0; (b) 0.48; (c) 16.8; (d) 48; (e) 86.4; (f) 288.

In the experiment, the chromatin can be stretched in two different ways (Brower-Toland *et al.,* 2002). One is to keep the stretching force constant (force clamp mode), which gives the time evolution of the DNA extension. The other is to keep the stretching velocity constant (velocity clamp mode), which gives the force-extension curve.

The simulated processes of nucleosome disruption under stretching in the force clamp mode with one and two nucleosomes are shown in Figs. 4 and 5, respectively. They look essentially just like the reverse of that in the cases of nucleosome formation (Figs. 2 and 3).

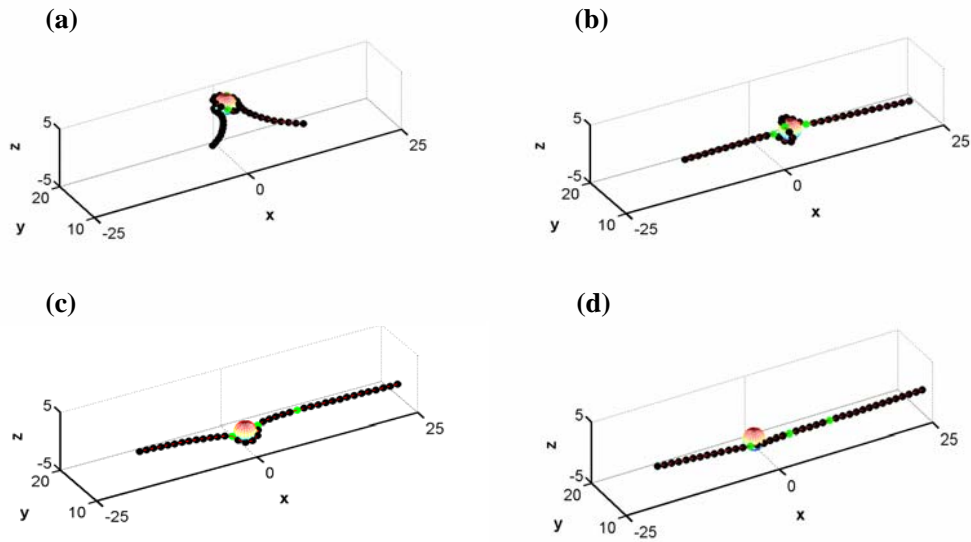

FIG 4. Snapshots of the dynamic process of a nucleosome disruption under stretching in the force clamp mode. The force applied is 8. The simulation time *t* is: (a) 0.5; (b) 3.8; (c) 5; (d) 5.9.

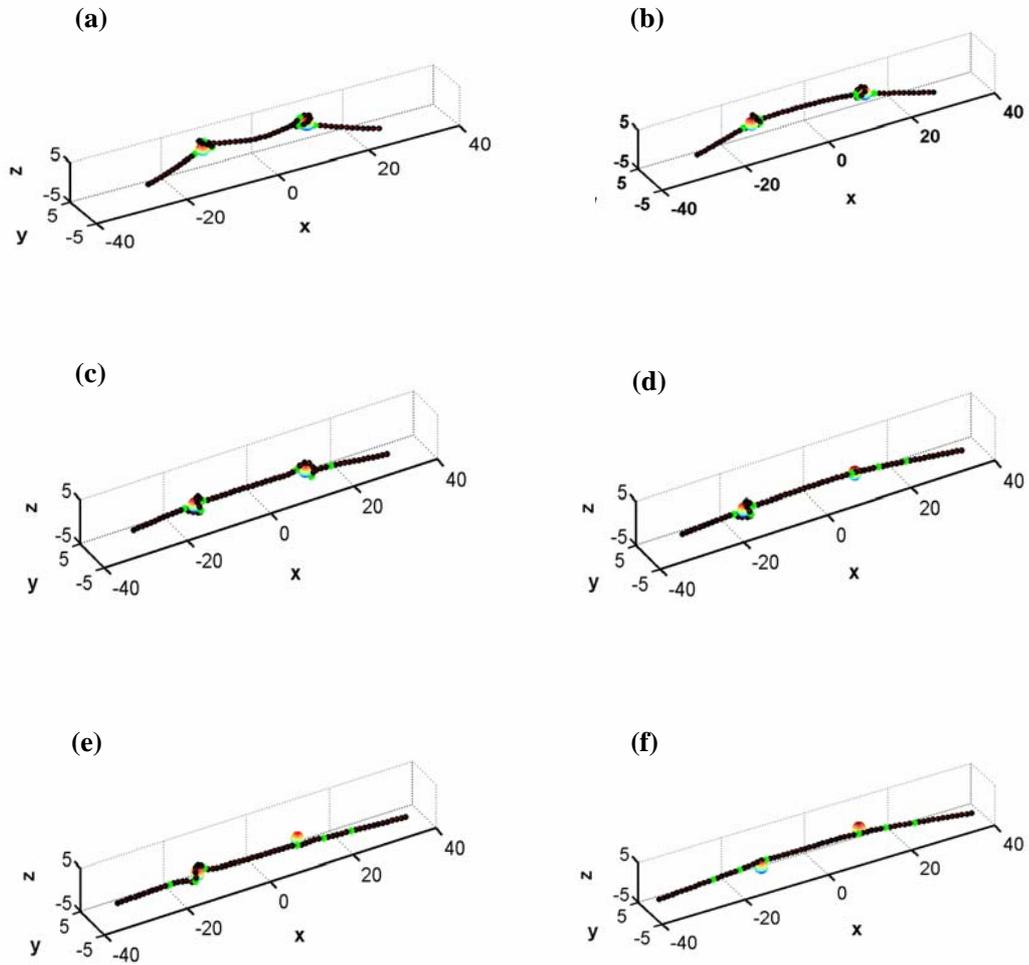

FIG 5. Snapshots of the dynamic process of two nucleosome disruptions under stretching in the force clamp mode. The force applied is 9. The simulation time $t$ is: (a) 1; (b) 3.8; (c) 5.6; (d) 8; (e) 9.6; (f) 11.

The extension vs. time and the force vs. extension curves for DNA with one nucleosome (two nucleosomes) are shown in Fig. 6 (Fig. 7). A nucleosome disruption corresponds to an obvious step in the extension vs. time curves, and to a peak in the force vs. extension curves. These results agree with the experimental observations (Brower-Toland *et al.,* 2002). More discussion will be given later.

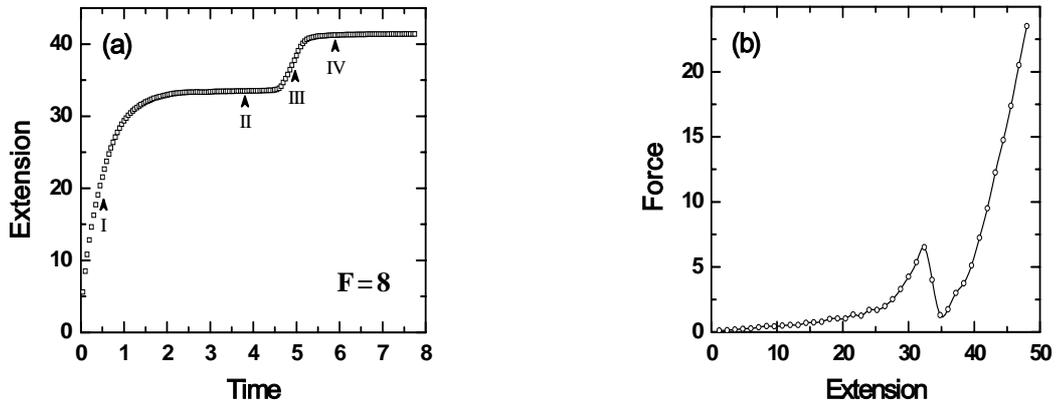

FIG 6. (a) Extension versus time curve of DNA with one nucleosome. *I* to *IV* correspond to the 4 snapshots in Fig. 4. (b) Force versus extension curve. A constant stretching velocity of 4.8 length unit per time unit is used.

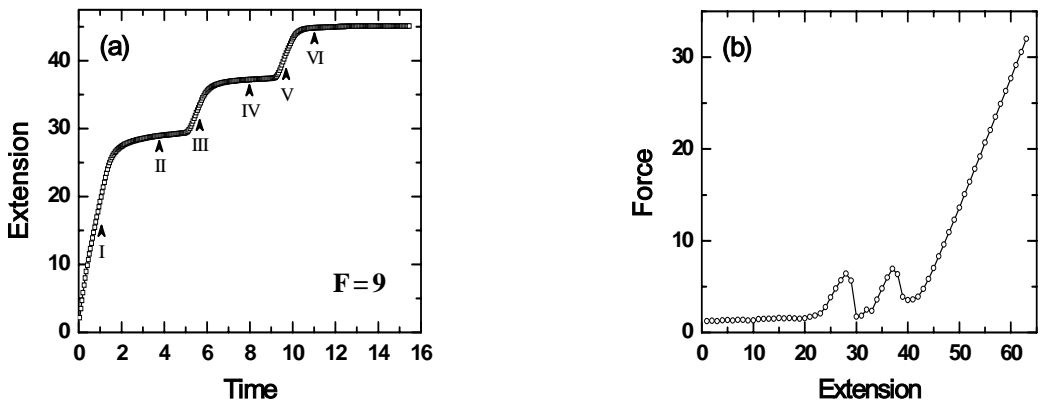

FIG 7. (a) Extension versus time curve of DNA with two nucleosomes. *I* to *VI* correspond to the 6 snapshots in Fig.5. (b) Force versus extension curve. A constant stretching velocity of 4.8 length unit per time unit is used.

## 4. Discussion

Our simulation results (Fig. 2 and Fig. 3) show an overall picture of the DNA wrapping process around the histone octamers. From these results we can see that the wrapping process almost involves only the DNA parts which are near the histone octamers, and the rest of DNA has no sharp structural changes. In addition, a histone octamer has some specific sites which strongly bind with DNA (Luger *et al.,* 1997; Luger and Richmond, 1998; Wolffe, 1998). Because of these fixed binding positions for DNA on its surface the histone octamer should rotate simultaneously in the DNA wrapping process. With these considerations, we propose an octamer-rotation model for the nucleosome formation process as shown in Fig. 8. Because in our model the histone octamer is taken as a spherical ball which has no any chiral priority, the modeled nucleosome may be either left-handed or right-handed (with equal possibilities). In nature, a histone octamer is consisted of 8 proteins and has a left-handed superhelix structure. Therefore, in the formation of a real nucleosome the histone octamer should rotate only in one direction determined by its chirality and the final nucleosome should be left-handed. For considering chirality in our future work, we will build up a more realistic model for the histone octamer. Although this will consume much more computer time for the simulation, we expect to obtain more information about the rotation of the histone octamers. The torsion of DNA will also need to be considered.

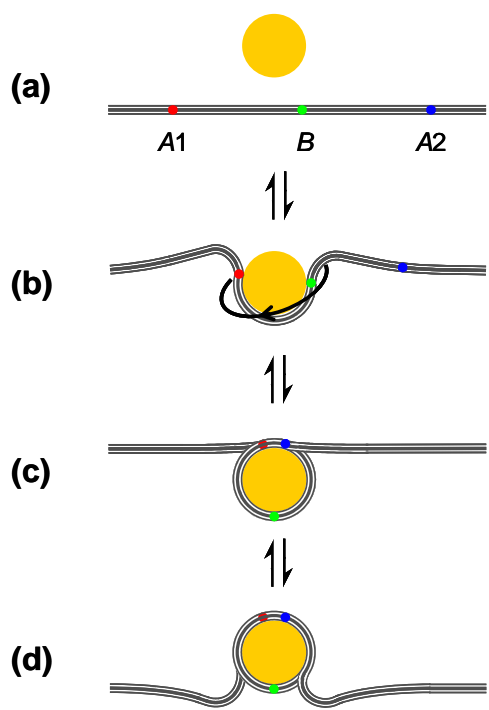

FIG 8. Octamer-rotation model for the process of nucleosome formation. The rotation direction of the histone octamer (with the binding DNA) determines the chirality of the nucleosome. In the present case, the histone octamer rotates in the clockwise direction and the chirality of the nucleosome is left-handed. If the histone octamer rotates in the anti-clockwise direction, the chirality of the nucleosome would be right-handed. Nucleosome disruption may be represented as from (d) to (a) [The rotation direction in (b) should be reversed in this case].

In the simulated stretching curves, we have sudden changes in extension or in force. Such a change corresponds to a separation between DNA and a histone octamer, that is, a nucleosome disruption. In the case of two nucleosomes there are two such changes, indicating that the nucleosome disruption occurs one by one. In the experiment of Brower-Toland *et al.,* (2002), sawtooth pattern was observed in force-extension curves of saturated nucleosomal arrays, also suggesting sequential nucleosome disruption. So whether some order exists in the disruption of individual nucleosomes or whether the process is purely stochastic (Hayes and Hansen, 2002)? We repeat our simulation with two nucleosomes and find they are disrupted randomly. In the future we will address this question by doing simulation work on stretching-induced disruption of more nucleosomes.

It is believed that a step in the extension vs. time curves (force clamp mode with high force) in experiment (Brower-Toland *et al.,* 2002) corresponds to a process from (c) to (b) to (a) in Fig. 8 [It should be noted that the rotation direction in (b) should be reversed for the stretching process]. Their measured step size was about 26 nm. In our preceding simulations, the distance between $A1$ and $A2$ is 12 length units (corresponding to 27.6 nm, or 81 bp). With this distance value, we obtain a theoretical step size of 8.5 length units (corresponding to 19.55 nm) which is shorter than the $A1$-$A2$ distance. This is because in our simulation, the strong DNA-histone-interaction locations lie on the DNA chain and during stretching the two S-locations are already separated to the opposite sides of the histone octamer sphere [see, for example, Fig. 4(b)] before the step occurs [Fig. 6(a), point II]. Actually, the strong DNA-histone-interaction locations lie on the histone octamer surface and thus the DNA loop remains closed on the histone octamer as shown in Fig. 8(c) before the step occurs. We also noticed that, for this $A1$-$A2$ distance (27.6 nm), our theoretical step size value (19.55 nm) is smaller than the experimental one (~26 nm) (Brower-Toland *et al.,* 2002). Because the step size is strongly dependent on the distance between $A1$ and $A2$, as shown in Fig.9, we can fit the experimental step size by adjusting $A1$-$A2$ distance. We find that with an $A1$-$A2$ distance value of 15 length units (34.5 nm), the simulation gives a step size of 26.45 nm. This $A1$-$A2$ distance corresponds to 101 bp of DNA which is very close to the 90-bp DNA length between the two strong binding sites of the histone octamer at superhelix locations ±4.5 (Luger and Richmond, 1998).

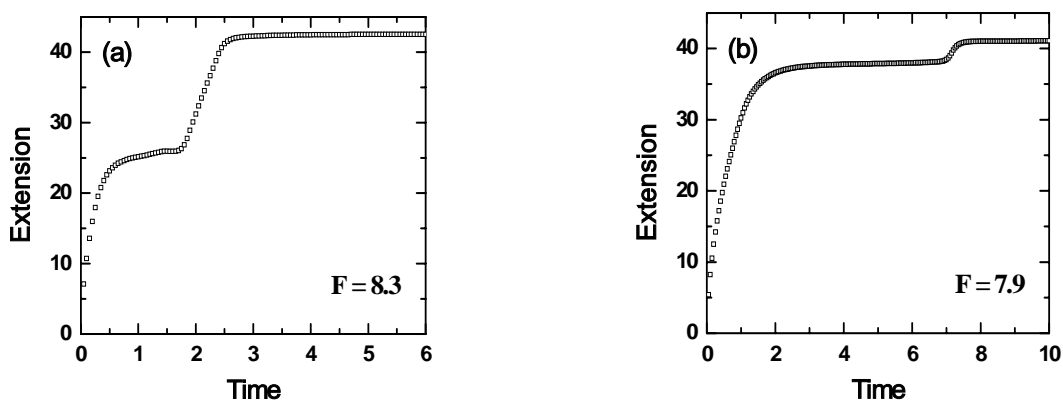

FIG 9.   Extension versus time curves with two $A1$-$A2$ distances: (a) 20 length units

and (b) 6 length units. The step sizes are 15 and 4 length units, respectively.

Here we discuss the effect of the relative binding strengths of the S-locations. In all the above simulations, the S-location *B* has been assumed to have weaker interaction with the histone octamer than the S-locations *A*1 and *A*2 [ $\varepsilon' = 15\varepsilon$ and $\varepsilon'' = 3\varepsilon$ in Eqs. 4(b)-(d)]. Assuming there is no S-location *B*, we find that there is no obvious change in the stretching curve, as shown in Fig. 10(a). On the contrary, when the interaction of the histone octamer with S-location *B* is as strong as that with *A*1 and *A*2, we find that the step is decomposed to two substeps [see Fig. 10(b)] due to the prolonged existence in the unwrapping process of the structure represented by Fig. 8(b). It should be noted that in the experiment of Brower-Toland *et al.*, (2002) such subteps have not been observed. One possible reason is that the substeps are absent because some allostery or cooperativity is operative and thus the unwrapping of the two halves of DNA in the nucleosome [see Fig. 8(c)] occurs simultaneously (Hayes and Hansen, 2002) or because the interaction of the histone octamer with the S-location *B* is indeed not strong enough. Another possible reason is that although there exist substeps, the time interval between the two substeps is relatively too short or the second substep is too small to be resolved in the experiment. We think the latter is more probable because firstly the two S-locations *A*1 and *A*2 bind to different histone contacts sites (Luger and Richmond, 1998) and therefore, the cooperativity, if existing, should be weak; and secondly, the structural data show that the S-location *B* has the highest binding stability (Luger and Richmond, 1998) and therefore, the interaction strength of the histone octamer with S-location *B* is at least as large as that with *A*1 and *A*2.

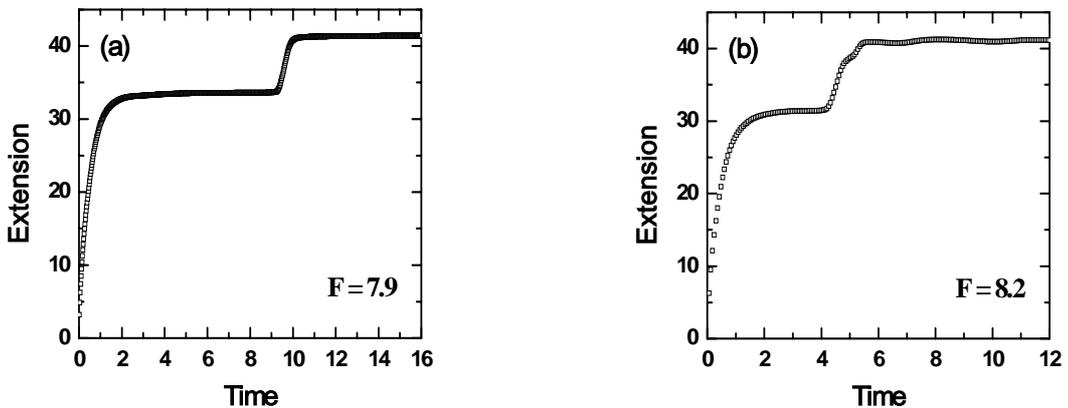

FIG 10. Extension versus time curves of DNA with one nucleosome. (a) The S-location *B* is absent; (b) The S-location *B* is as strong as *A*1 and *A*2 when interacting with a histone octamer.

We also simulated the stretching process for DNA without any S-location. As shown in Fig. 11, both the step in the extension vs. time curve and the peak in the force vs. extension curve disappear.

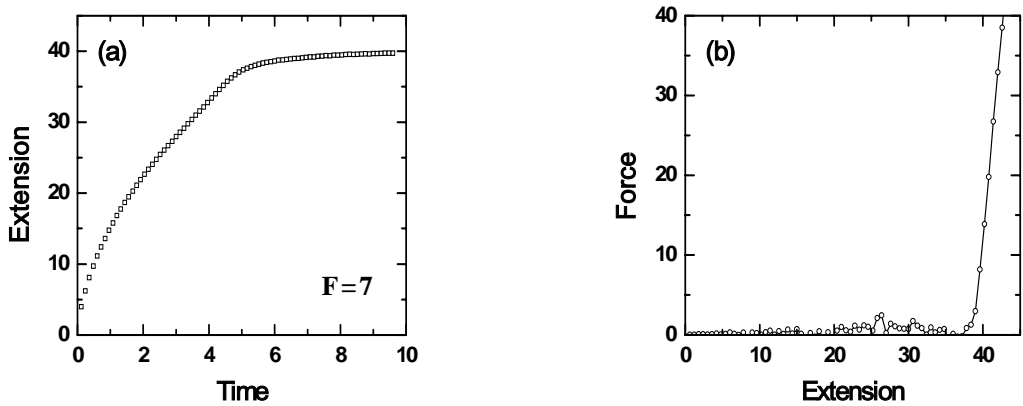

FIG 11.  Stretching curves of a DNA with one nucleosome. There is no S-locations on the DNA. (a) Extension versus time curve; (b) Force versus extension curve.

## 5. Summary

We have studied the processes of nucleosome formation and disruption by Brownian dynamics numerical simulation. Locations where DNA interacts strongly with histone octamers are included. From the simulation results, we propose an octamer-rotation model to describe the process of interaction between DNA and histone octamer(s) during nucleosome formation. In the present case of spherical histone octamers, it is the direction of octamer rotation that determines the chirality of the nucleosome formed. Actually, histone octamers are left-handed and this chirality should determine the octamer-rotation direction as well as the chirality of the nucleosome formed.

The disruption process of a nucleosome structure under stretching has also been simulated. From simulations we get stretching curves that are consistent with previous experimental results. The appearance of steps in the extension vs. time curve (or peaks in the force vs. extension curve) demonstrates that the interaction strength of DNA with histone octamers is indeed not uniform in the nucleosomes.

In our future work, to get a more precise picture of the interaction process of DNA and histone octamers, we need to make an improved model for the histone octamer that includes spatial distribution of its strong DNA-binding sites. The torsion of DNA chain will also be taken into account. We will also investigate how individual nucleosomes are disrupted sequentially when there are more nucleosomes.

**Acknowledgements**

This research was supported by National Natural Science Foundation of China Grants 60025516 and 10334100, and the Innovation Project of the Chinese Academy of Sciences.